\documentclass[article,twocolumn,
amsmath,amssymb,aps,prl]{revtex4}

\usepackage{color}
\usepackage{graphicx}
\usepackage{dcolumn}
\usepackage{braket}
\usepackage{bm}
\usepackage{amsmath,amssymb,amsthm}
\usepackage{lipsum} 
\usepackage{hyperref}
\usepackage{comment}
\usepackage{braket}

\usepackage[T1]{fontenc}

\begin{document}

\author{Filippo Caleca}
\affiliation{%
ENS de Lyon, Universit\'e Lyon 1, CNRS, Laboratoire de Physique, F-69342 Lyon, France
}%
\author{Fabio Mezzacapo}
\affiliation{%
ENS de Lyon, Universit\'e Lyon 1, CNRS, Laboratoire de Physique, F-69342 Lyon, France
}%
\author{Tommaso Roscilde}
\affiliation{%
ENS de Lyon, Universit\'e Lyon 1, CNRS, Laboratoire de Physique, F-69342 Lyon, France
}%

\title{Breakdown of thermalization in quantum spin systems with long-range interactions}

\begin{abstract}
Quantum thermalization of closed systems -- i.e. the relaxation of local properties to their thermodynamic value along the Hamiltonian evolution -- is generically expected in non-integrable systems at finite energy density. In this work we show that quantum spin systems with at least one spin component coupled through long-range ferromagnetic interactions, and a second spin component coupled to a field, cannot in general thermalize to a paramagnetic state whenever they have a finite, negative energy per spin. The long-range interactions we consider decay as a power law of the distance, $\sim r^{-\alpha}$ with $\alpha < d$, where $d$ is the number of spatial dimensions of the system. The lack of thermalization is generic to an infinite family of Hamiltonians, and it is not associated to any form of integrability. It can instead be understood by the anomalous features of the thermodynamics of the system in the paramagnetic phase, which cannot be reflected in the properties of the  eigenstates. This leads to a breakdown of eigenstate typicality and of the eigenstate thermalization hypothesis. Interestingly, the non-thermal states emerging at long times in the dynamics display long-range multipartite entanglement, elevating the breakdown of thermalization to a resource.   
\end{abstract}

\maketitle

\emph{Introduction.}
The dynamics of closed quantum many-body systems away from equilibrium and their relaxation towards equilibrium 
\cite{Polkovnikov2011,dalessio_quantum_2016,AletL2018,Abaninetal2019} are central subjects of modern quantum statistical physics. Major theoretical advances have shown that, for generic quantum systems with a Hamiltonian $H$, and for generic initial states $|\psi_0\rangle$, the unitary dynamics can realize thermal equilibrium at the level of local observables, owing to fundamental properties of \emph{typicality} of the Hamiltonian eigenstates $\{ |\psi_a\rangle \}$. 
By thermalization one generally means that, in a system of size $N$, the long-time averages of local observables $O$ (e.g., on-site observables, two-point correlations, etc.)
tend to reproduce the averages over the Gibbs ensemble, i.e. $\frac{1}{T} \int_{t_0}^{t_0+T} dt~ \langle \psi(t)| O | \psi(t) \rangle \approx  {\rm Tr} (O e^{-\beta H})/  {\rm Tr}(e^{-\beta H}) = \langle O \rangle_\beta$. Here, 
$t_0$ is some characteristic thermalization time; $T$ is the averaging time, taken to be much larger than the time scale for microscopic dynamics;   $|\psi(t)\rangle = e^{-iHt} |\psi_0\rangle$ (with $\hbar=1$) is the solution of Schr\"odinger's equation; and $\beta$ is the inverse temperature associated with the initial state, such that $E_0 = \langle \psi_0 | H |\psi_0\rangle =: {\rm Tr} ( e^{-\beta H} H ) / {\rm Tr}(e^{-\beta H})$. If the dynamics conserves a finite number of global quantities $Q_k$ beyond the energy (e.g., the particle number, the magnetization, etc.), then the Gibbs ensemble should be modified $e^{-\beta H} \to e^{-\beta H - \sum_k \lambda_k Q_k}$, with $\lambda_k$ acting as Lagrange 
multipliers that fix the average value of the conserved quantities.  
The physical, intuitive picture underlying local thermalization is that, for local observables related to a subsystem $A$ with size $\ll N$, the complement subsystem $B$ can be viewed as a reservoir, inducing canonical equilibrium locally. 
The eigenstate thermalization hypothesis (ETH) \cite{Srednicki1999, Deutsch2018,dalessio_quantum_2016} provides a sufficient condition for local thermalization to occur, requesting that most eigenstates close to the energy $E_0$ possess eigenstate expectation values (EEVs) which reproduce the Gibbs-ensemble average. In mathematical terms, $O_{aa} = \langle \psi_a | O |\psi_a \rangle = O(E_a) + {\cal O}(e^{-S_{E_a}/2})$ \cite{Srednicki1999, Deutsch2018} with $S_{E_a} \sim N$ the microcanonical entropy at the eigenstate energy $E_a$, implying that EEVs are smooth functions of the energy up to corrections decreasing exponentially with $N$, and $O(E_0) = \langle O \rangle_\beta$.  

The above assumption has been numerically verified in a variety of different systems \cite{Beugelingetal2014,dalessio_quantum_2016}. Yet various mechanisms have been identified as possibly leading to the breakdown of the ETH; and they have been shown to result in a failure of the system's dynamics to thermalize (as defined before). A general principle for ETH breakdown is the existence of a large number of conserved quantities beyond energy, as realized in integrable systems \cite{Essler2016,Vidmar2016,Vasseur2016}; and as conjectured in many-body localized systems \cite{AletL2018,Abaninetal2019}. Another important, alternative mechanism is Hilbert-space fragmentation \cite{Moudgalya2022}, namely the existence of a large number of sectors in Hilbert space which cannot be connected by the Hamiltonian. In all these examples, the dynamics of the system is constrained by much more than the energy conservation, i.e. either by the conservation of integrals of motion, or by the conservation of the Hilbert-space sector of the initial state. 
Important experiments on quantum simulators have verified the breakdown of thermalization in all these scenarios \cite{Kinoshita2006,Schreiberetal2015,Choietal2016,Kohlertetal2023,Hondaetal2025}.

\begin{figure}[ht!]
\begin{center}
\includegraphics[width=0.9\columnwidth]{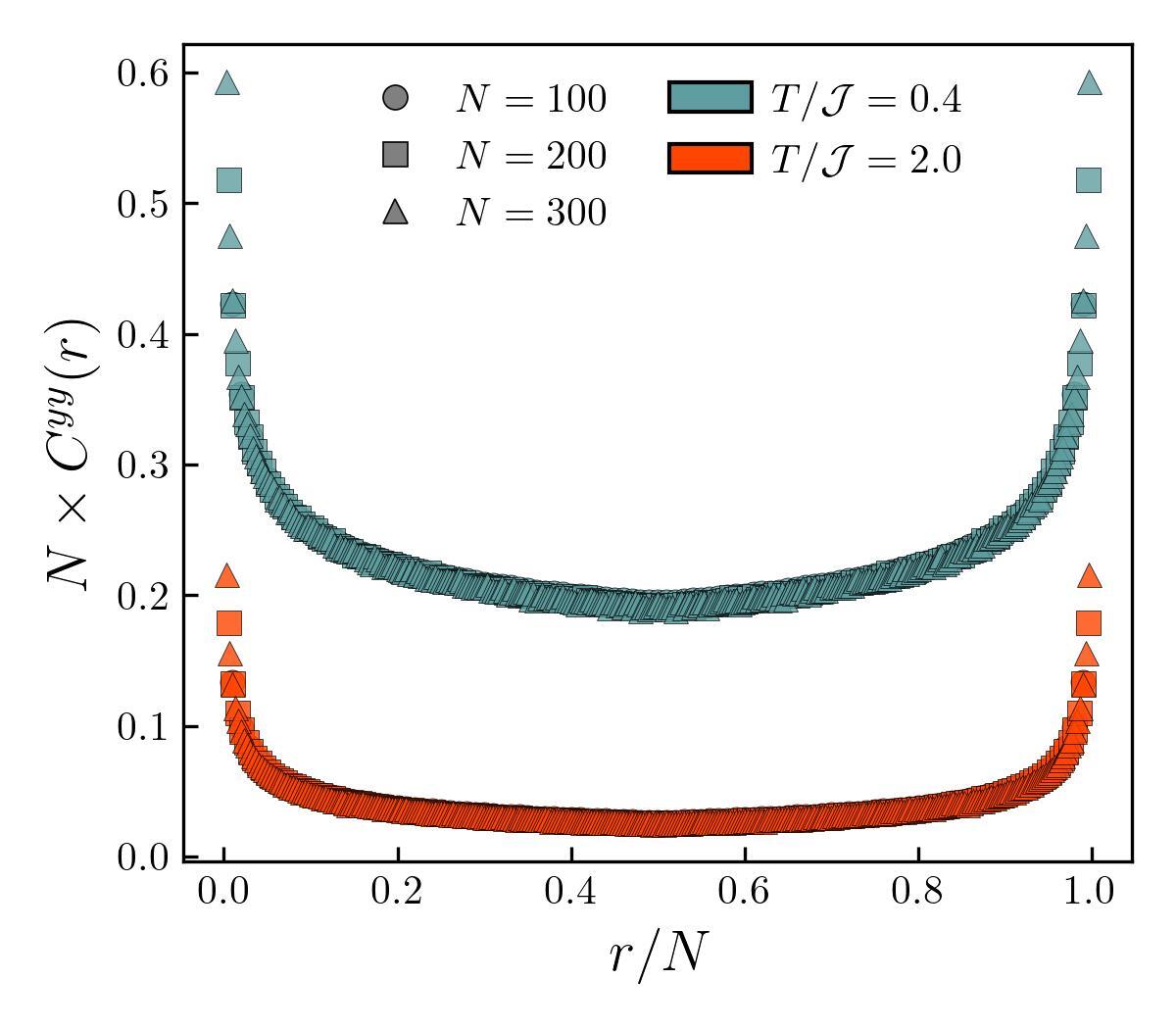}
\caption{\emph{Scale invariance of correlations in the paramagnetic phase.} Correlator $C_y(r)$ at two different temperatures ($T/J = 0.4$ and 2) and three different sizes ($N=100, 200,$ and 300) for the 1d LR XX model with $\alpha=0.5$ and a field $\Omega/{\cal J} = 0.3$. The curves for different sizes, obtained via QMC, collapse when plotted vs $r/N$ and multiplied by the system size $N$.}
\label{f.correlations}
\end{center}
\end{figure}

\begin{figure*}[ht!]
\begin{center}
\includegraphics[width=0.9\textwidth]{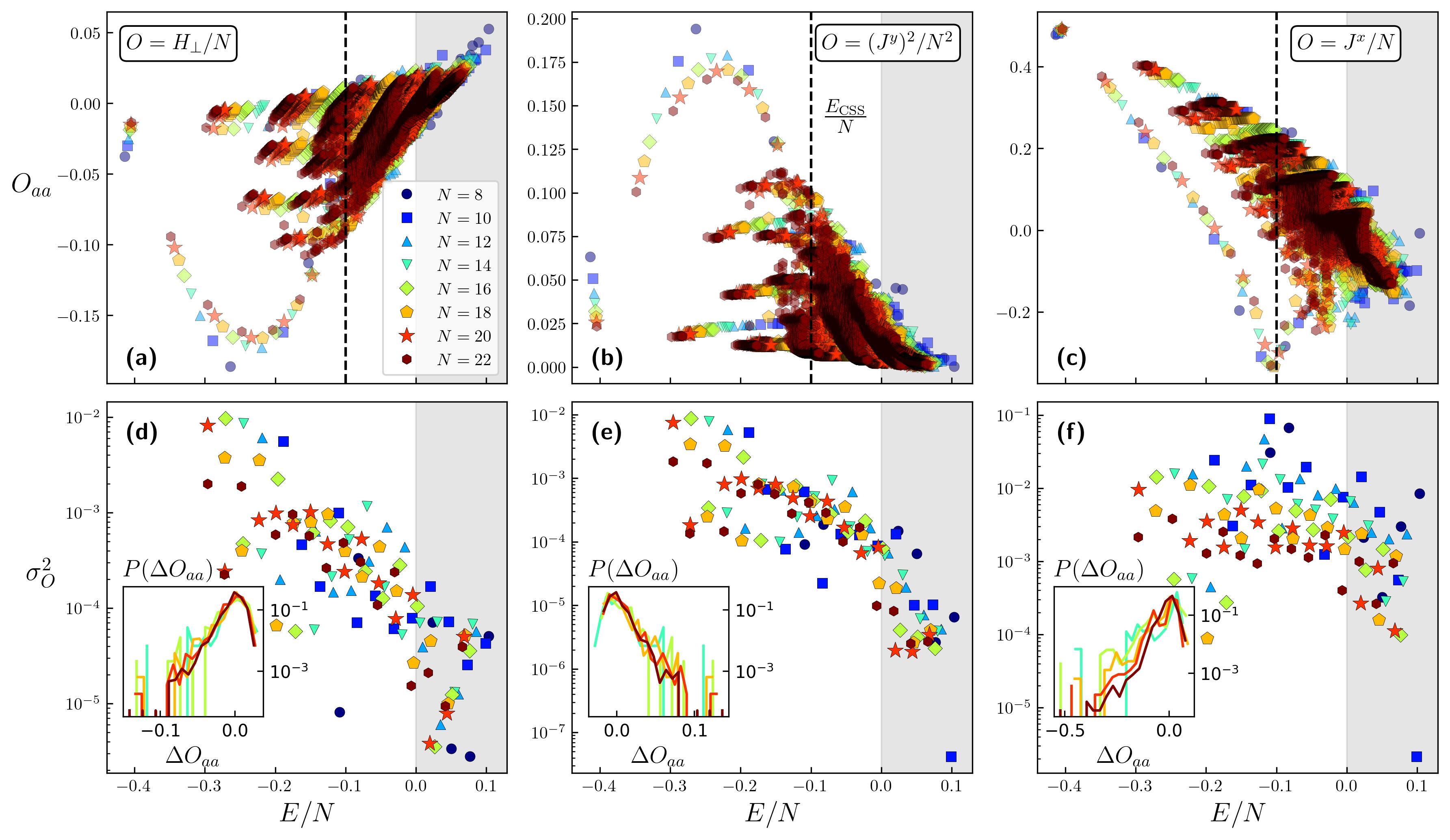}
\caption{\emph{Atypicality of the eigenstates with LR interactions.} (a, b, c) Eigenstate expectation values (EEV) $O_aa$ for three different observables, $O = H_{\perp}/N$, $(J^y)^2/N^2$ and $J^x/N$, for 8 different sizes, obtained for the 1d LR XX model with $\alpha=0.5$ and a field $\Omega/{\cal J} = 0.3$. The dashed line marks the energy corresponding to the coherent spin state (CSS) oriented opposite to the field.  (d, e, f) Scaling behavior of the mean squared deviations of EEVs, $\Delta O_{aa}$, from their local microcanonical average, for the three above observables. Insets: distribution of the local deviations of EEVs, $\Delta O_{aa}$, evaluated over the states with negative energy.}
\label{f.atypicality}
\end{center}
\end{figure*}

In this work we propose a new scenario for the breakdown of ETH and of thermalization, based on the presence of \emph{long-range (LR) interactions}. For definiteness, we shall focus on ensembles of $S=1/2$ spins cast on a $d$-dimensional lattice, with Hamiltonian   $H = \sum_{\mu=x,y,z} H_\mu + H_{\Omega}$, where
\begin{equation}
H_{\mu} = - \frac{1}{K_\alpha} \sum_{i\neq j}  \frac{{\cal J}^\mu}{r_{ij}^\alpha} S_i^\mu S_j^\mu ~~~~~ H_{\Omega} =  - \Omega \sum_i S_i^x  ~.
\label{e.Ham}
\end{equation}  
Here $i,j$ run on a lattice with $N$ sites, ${\cal J}^\mu > 0$ are the coupling constants of the various spin components, which we take to be all ferromagnetic to avoid any frustration effect;  $\Omega$ is an external field, which we take to be coupled to the $x$ spin component. Interactions are long-ranged when $\alpha < d$ \cite{Campa-book}: in the following we shall mostly consider systems with $\alpha <1  = d$. $K_\alpha = \sum_{\bm r \neq 0} r^{-\alpha} \sim N^{1-\alpha/d}$ is the Kac normalization factor, guaranteeing that the energy is extensive.
When $\alpha<d$, even upon Kac normalization the energy remains non-additive \cite{Campa-book}, because the system cannot be decomposed into subsystems whose internal energy is much larger than the subsystem-to-subsystem coupling energy. This aspect prevents fundamentally the notion that a system can act as its own reservoir, as already pointed out in the context of classical systems with long-range interactions \cite{Levinetal2014}. 
Indeed such systems, as \emph{e.g.}, gravitational systems in $d=3$, can fail to thermalize over very long time scales. Yet, in the classical context, the conditions for a system to be ergodic are not as precisely stated as in quantum mechanics. For quantum systems we can explicitly show that ETH is not verified by a very large class of models belonging to the family of Eq.~\eqref{e.Ham}, and hence establish firmly the absence of thermalization. We do so by showing that the thermodynamics of such models can have anomalous traits, which Hamiltonian eigenstates cannot generally verify. Under these conditions, eigenstates can display a systematic atypicality, showing much stronger correlations than in thermal equilibrium. The consequence of eigenstate atypicality is that dynamics that should relax to a paramagnetic phase if thermalization occured, leads instead to the onset of long-range correlations and entanglement. Hence the breakdown of thermalization in long-range interacting systems represents a quantum resource. 

\emph{Anomalous thermodynamics of LR interacting spin models.}
We reorganize the Hamiltonian into two terms, $H = H_{||} + H_{\perp}$, where $H_{||} = H_{\mu=x} + H_{\Omega}$ contains all the terms diagonal in the $S^x$ eigenbasis; while $H_{\perp} = \sum_{\mu=y,z} H_\mu$ contains the off-diagonal terms in that basis.
A feature of quantum states which is crucial for the following analysis is their correlations, specifically the long-distance behavior of the correlation functions $C_{\mu}(\bm r) =  \langle S_i^\mu S_{i+\bm r}^\mu \rangle$, where $i+{\bm r}$ is the site at a distance ${\bm r}$ from site $i$. Long-range order for the $\mu$ spin component corresponds to $\lim_{|\bm r|\to\infty} C_{\mu}(\bm r) = m^2_\mu \neq 0$.
For all models described by Eq.~\eqref{e.Ham}, a sufficiently strong field $\Omega$ leads the system to a paramagnetic phase at equilibrium, \emph{i.e.}, a phase for which all correlations for spin components transverse to the field are short-ranged, such that $m_y = m_z = 0$. In the following we focus on this paramagnetic phase, as it possesses a remarkably anomalous property, at the root of the breakdown of ETH.  
It is easy to realize that the Hamiltonian term $H_\mu$ in Eq.~\eqref{e.Ham} has an extensive expectation value $\langle H_{\mu} \rangle \sim O(N)$ if and only if the state of the system displays long-range correlations for the $\mu$ spin component, \emph{i.e.}, iff $m_\mu \neq 0$. Indeed, under this condition one has that $\langle H_{\mu} \rangle \approx J^{\mu} m^2_\mu N$. Otherwise $\langle H_{\mu} \rangle$ becomes sub-extensive, because of the presence of the Kac normalization factor. As an example, for correlations decaying as a power law $C_{\mu}(\bm r) \sim r^{-\rho}$ ($\rho>0$), one has $\langle H_{\mu} \rangle/N \sim K_\alpha^{-1} \int_a^{N^{1/d}} dr ~r^{d-\rho-\alpha-1} \sim N^{-\rho/d}$ (where $a$ is the lattice spacing). 

We now focus on the paramagnetic phase, and introduce the collective-spin operator ${\bm J} = \sum_i {\bm S}_i$. In this phase we have that $\langle H_{||} \rangle \sim O(N)$ by virtue of the field term, since the magnetization is extensive, \emph{i.e.}, $\langle J^x \rangle \sim O(N)$. On the other hand, by definition of paramagnetism, $\langle H_{\perp} \rangle/N \to 0$ when $N\to \infty$. This implies that the Hamiltonian term $H_{\perp}$, in spite of having a spectral width scaling in the same way as that of $\langle H_{||} \rangle$ (by virtue of Kac normalization), disappears in practice from the energetics of the system as $N$ increases. This anomalous property results in the atypicality of Hamiltonian eigenstates. Consider the observable $O = H_{\perp}/N$. In order to satisfy ETH, one should have that the EEV $O_{aa}$ conforms with the equilibrium value as $N\to \infty$, namely $O_{aa}$ should become vanishingly small, while the eigenenergy $H_{aa}/N = E_a/N$ should generally remain finite. This would imply that, in order to be typical and conform with ETH, eigenstates $|\psi_a\rangle$ of $H = H_{||} + H_{\perp}$  should approach eigenstates of $H_{||}$ when $N\gg 1$, in spite of the fact that $[H_{||}, H_{\perp}] \neq 0$ -- since, if $O_{aa} \to 0$, then $H_{aa} = E_a \approx (H_{||})_{aa}$.  This is clearly a paradox, implying that ETH cannot be satisfied in the paramagnetic phase. 
 
We verify the anomalous property of the paramagnetic thermodynamics for the case of an XY model, i.e. ${\cal J}^x = {\cal J}^y ={\cal J}$, ${\cal J}^z = 0$ with $\alpha = 0.5$ in $d=1$ -- such an Hamiltonian with $\alpha<1$ can be realized \emph{e.g.}, with trapped ions \cite{Richerme2014, Jurcevic2014, Brydges2019}. This model is in a paramagnetic phase for any $\Omega \neq 0$, since the field couples to the order parameter ($J^x$) of the system. A quantum Monte Carlo (QMC) calculation \cite{Sandvik2010AIPCP} of the correlation properties for $\Omega/{\cal J}=0.1$ (Fig.~\ref{f.correlations}) shows that $C^{y}$ approximately satisfies the scaling law $C^{y}(r) \approx N^{-1} g(r/N)$, which implies $\langle H_{\perp} \rangle \sim -{\cal J} \int dx ~g(x) x^{-\alpha} \sim -O(1)$ (provided that the integral converges).  Therefore the Hamiltonian term $H_{\perp}$ disappears from the thermodynamic behavior for large systems at any temperature $T<\infty$, namely when $|\langle H \rangle| \sim O(N)$. Our QMC calculation is made in the canonical ensemble, while the ensemble relevant for closed-system thermalization is the microcanonical (MC) one, and the two ensembles can be inequivalent in the case of long-range interactions \cite{Campa-book}. Nonetheless for $N\to \infty$ this only happens in the vicinity of first-order phase transitions \cite{Touchetteetal2004, Bouchet2005}, which are absent in the paramagnetic regime. Hence our QMC calculations are also relevant for the MC behavior when $N \gg 1$. {Further discussion about ensemble equivalence is provided in the Supplemental Material (SM) \cite{SM}. }


\begin{figure*}[ht!]
\begin{center}
\includegraphics[width=0.85\textwidth]{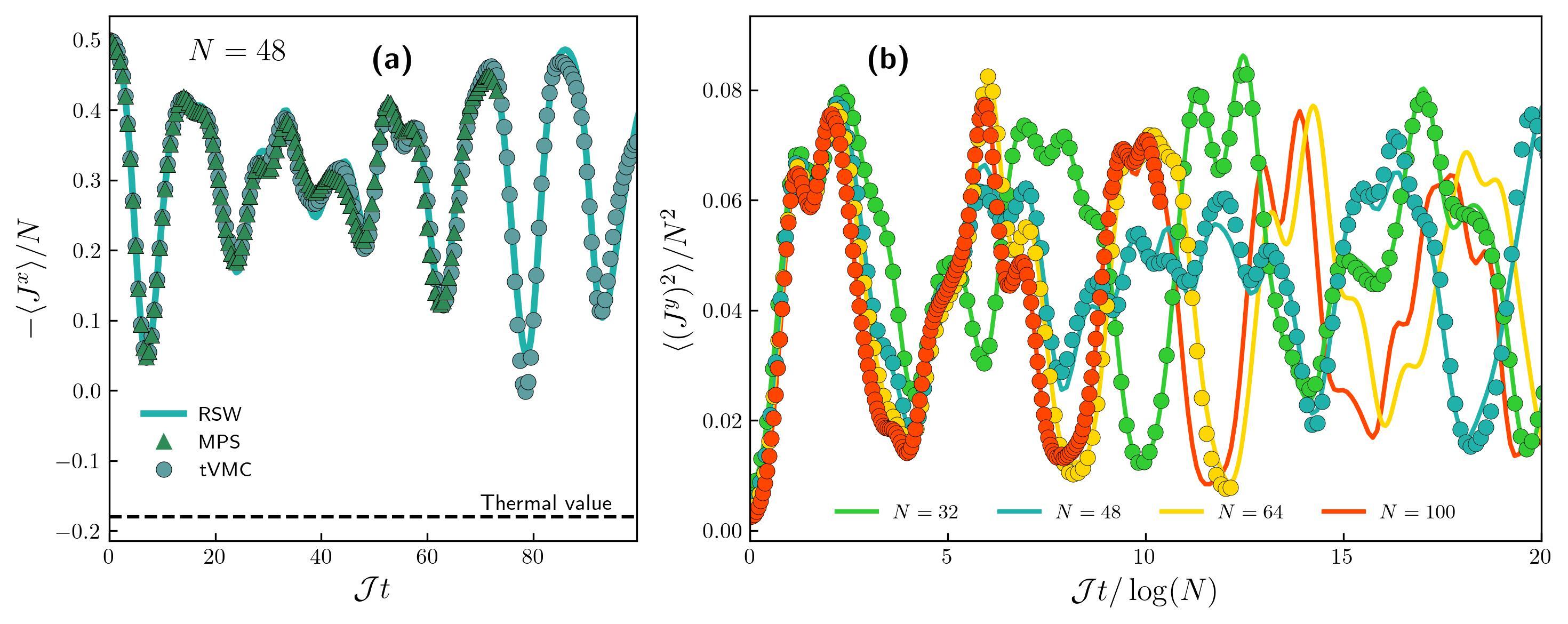}
\caption{\emph{Non-thermalizing dynamics.} (a) Dynamics of the average magnetization per spin $\langle J^x \rangle/N$ in the 1d LR XX model with $\alpha=0.5$ and a field $\Omega/{\cal J} = 0.3$, starting from the coherent spin state opposed to the field. The data refer to rotor+spin-wave (RSW) calculations, matrix-product state (MPS) and time-dependent variational Monte Carlo (tVMC) calculations for $N=48$. The dashed line marks the expected thermal value. (b) Same as in (b) but for  $\langle (J^y)^2 \rangle/N^2$ and various system sizes. Rescaling by $\log N$ of the time axis allows for the alignment of the first peak, as expected in TaT dynamics \cite{Roscilde2025}.}
\label{f.dynamics}
\end{center}
\end{figure*}

\emph{Atypicality of Hamiltonian eigenstates.} Having established that typicality of eigenstates in the paramagnetic phase would lead to a paradox, we verify their atypical nature. We focus on the same long-range XY model in $d=1$, for various values of $\Omega/\cal J$, and perform exact diagonalization (ED) of the Hamiltonian up to $N=22$ spins in the zero-momentum sector, which, beside reducing the Hilbert space dimension, is also the one relevant to the dynamics that we investigate in the next section. First of all, energy levels exhibit repulsion, and a distribution of level spacings rather close to Wigner-Dyson statistics (see Methods), confirming that the model of interest does not display any obvious form of integrability. Nonetheless the distribution of EEVs expose the atypicality of eigenstates. Figs.~\ref{f.atypicality}(a,b,c) shows that the EEVs of the observables $H_{\perp}/N$, $(J^y)^2/N^2$ and $\langle J^x \rangle/N$  
are widely spread, and do not collapse onto a smooth $O(E)$ curve upon increasing the system size, but they rather maintain their spreading, \emph{i.e.}, their atypicality. This is also observed in half-system entanglement entropies, as shown in the Methods section. To assess this aspect quantitatively, following Ref.~\cite{Beugelingetal2014} we divide the axis of the energy per spin $E/N$ into $m=20$ equal bins. For each state in the bin we define the deviation of its EEV from the local MC average $\Delta O_{aa} = O_{aa}- \bar O(E_a)$,  where $\bar O(E_a)$ was obtained by averaging $O_{aa}$ over the eigenstates in an energy window $[E_a- \delta E, E_a+\delta E]$ with $\delta E = W/m$ ($W$ being the full spectral width); and we evaluate the average fluctuations around the local MC averages $(\sigma_O^2)_b =  N_b^{-1} \sum_{a \in b} (\Delta O_{aa})^2$, where $N_b$ the number of states in bin $b$.  According to ETH, $(\sigma_O^2)_b$ should decrease \emph{exponentially} with system size, or polynomially with the number of states of the system -- in particular Ref.~\cite{Beugelingetal2014} finds that $(\sigma_O^2)_b \sim N_b^{-1}$; and this behavior is indeed exhibited by the model of our interest in the case of short-range interactions ($\alpha=3$) as shown in the Methods section. On the other hand, Figs.~\ref{f.atypicality}(d,e,f) show that for long-range interactions, $\alpha=0.5$, $(\sigma_O^2)_b$ does not seem to scale at all, at least in the regime of negative energies of our interest, $E/N<0$, which corresponds to the physics of ferromagnetic long-range interacting models. Some scaling appears for $O=J_x/N$, but it is absent for a different field value ($\Omega/{\cal J}=0.1$), as shown in the Supplemental Material (SM) \cite{SM}. Accordingly, the distribution $P(\Delta O_{aa})$ of EEV-to-MC-average deviations does not significantly shrink with system size, and it has a highly non-Gaussian form; whereas it shrinks exponentially in the case of short-range interactions (see Methods) \cite{Beugelingetal2014}. 
Interestingly, atypicality of eigenstates can come with very clear physical consequences: atypical states can have $(H_{\perp})_{aa}  \sim O(N)$, which can only happen in the presence of long-range transverse correlations, $m^y\neq 0$, resulting in $[(J^y)^2]_{aa} \sim O(N^2)$; and they accordingly have anomalously low entanglement entropies. The significant contribution of  $(H_{\perp})_{aa}$ to the total energy results in large deviations of $(H_{||})_{aa}$ from the thermal value, which can come from the magnetization in the term $-\Omega J^x$  deviating from its thermal value, including the possibility that the magnetization is \emph{opposite} to the applied field. In the Methods section we estimate the number of atypical states to grow exponentially with system size, while remaining a set of zero measure within their symmetry sector.
Their distinct physical traits not only dominate the fluctuations of the EEVs, but they also have a direct consequence for the dynamics of the system, as discussed below.  

\emph{Non-thermalizing quench dynamics leading to long-range correlations.} The violation of ETH in the properties of the Hamiltonian eigenstates should lead to non-thermalizing dynamics, i.e., producing states whose properties can deviate systematically from what expected at thermal equilibrium, at the energy $E_0$ of the initial state. In ETH-violating systems such as many-body localization or Hilbert-state fragmentation, lack of thermalization is generally associated with persisting memory of the initial state in the evolved state: the initial state overlaps primarily with atypical Hamiltonian eigenstates, that therefore dominate the ensuing dynamics. A similar behavior can be expected in our proposed scenario of ETH violation by long-range interactions, but with an extra twist: atypical states deviate from the thermodynamic behavior in their \emph{scaling} properties (of energies and of correlations), so that during its dynamics the system can produce new many-body \emph{phases of matter} compared to what is possible at equilibrium. In this work we focus on the case of a natural initial state for experiments, \emph{i.e.}, a coherent spin state anti-aligned with the field $|\psi(0)\rangle = |{\rm CSS}_{-x}\rangle =  \otimes_{i=1}^N  |\leftarrow_x\rangle_i$. This state maximizes the energy of interaction with the field, which, for sufficiently large $\Omega$, positions it in the bulk of the Hamiltonian spectrum (see Fig.~\ref{f.atypicality} for $\Omega/{\cal J} = 0.3$). Having maximally negative magnetization, the state can overlap strongly with eigenstates which have anomalously large and negative magnetization, as well as long-range $C_y$ correlations. 

We investigated the dynamics of the 1d LR XX model with $\alpha=0.5$ initialized in the CSS$_{-x}$ by making use of two Ansatz-based approaches which allow us to far exceed the sizes accessible to ED: 1) matrix-product states (MPS) up to sizes $N=48$ (making use of the ITensor library \cite{itensor}); and 2) time-dependent variational Monte Carlo (tVMC) up to $N=100$ (making use of a pair-product Ansatz~\cite{PRB2019, Comparin2022PRA}, generalized by including a magnetization-dependent term \cite{PRL2025FC,Roscilde2025}). 
All of our results -- see Fig.~\ref{f.dynamics} for $\Omega/{\cal J}=0.3$ -- show that the magnetization remains persistently opposite to the field. Concomitantly long-range correlations develop in the system, leading to $\langle (J^y)^2 \rangle \sim O(N^2)$ -- and this in spite of the fact that the thermodynamic phase is paramagnetic, \emph{i.e.}, $\langle (J^y)^2 \rangle \sim O(N)$, and with an average magnetization opposite to the dynamical one. Hence the striking properties of atypical states become fully visible in the dynamics. In fact the dynamics of the system can be understood in connection with that of the all-to-all interacting system, \emph{i.e.}, the $\alpha=0$ limit, known as twist-and-turn (TaT) dynamics \cite{Michelietal2003,Muesseletal2015,MunozArias2023,Roscilde2025}. Rotor+spin-wave (RSW) theory \cite{Roscildeetal2023b,Roscildeetal2023,Roscilde2025,SM} treats the effect of a finite value of $\alpha$ as the appearance of harmonic spin-wave fluctuations on top of the collective-spin (or rotor) dynamics, and decoupled from the latter, as long as the spin-wave excitations are dilute. This implies in turn that the system should exhibit persistent TaT dynamics of the rotor, without the latter ever equilibrating with the spin waves. The quantitative agreement between the RSW prediction and the numerical results, shown in Fig.~\ref{f.dynamics}, validates the RSW picture as emerging when thermalization breaks down. In fact, in the SM \cite{SM} we show that RSW becomes increasingly accurate as the size is increased, since the density of spin-wave excitations generated in the dynamics decreases to zero. 

\emph{Discussion and conclusions.} Several previous works have pointed out that long-range interactions in quantum spin models can lead to possible deviations of the eigenstate properties from ETH \cite{Russomannoetal2021,  Sugimotoetal2022}, to the presence of quantum many-body scars \cite{Leroseetal2025,CorpsR2026}; and a generalization of ETH for long-range interactions has been proposed \cite{PalS2025, Srirametal2026}. Our work shows explicitly that atypicality, contradicting ETH, is not just a characteristic of a small fraction ($O({\rm poly}(N))$) of states in the spectrum, but it is actually widespread, affecting $O(\exp(N))$ states; and that its roots can be found in the anomalous nature of the thermal equilibrium behavior. Atypicality is so strong as to fundamentally dominate the scaling behavior of fluctuations of eigenstate expectation values around the microcanonical averages, leading to a breakdown of ETH. The form of atypicality manifested by long-range systems is rather drastic, as it manifests itself in the very scaling behavior of correlations properties. It is shown to lead to closed-system dynamics which does not simply fail to thermalize, but which produces in fact a totally different phase of matter compared to equilibrium. In particular, the phase of matter emerging dynamically exhibits long-range quantum correlations, which are associated with an extensive quantum Fisher information density (corresponding \emph{e.g.}, to $4\langle (J^y)^2 \rangle/N$ in our calculations).  These correlations are accompanied by the conservation of the parity of the $J^x$ collective spin component, and, as shown in Ref.~\cite{FrerotR2024}, they have an immediate metrological significance, as they can be optimally exploited for phase estimation by measuring the evolution of the parity upon application of the unitary transformation $e^{-i J^y \theta}$.  Therefore the breakdown of thermalization in long-range interacting systems surges to the level of a quantum resource.   
Our observations are relevant to several quantum simulation platforms with long-range interactions, such as trapped ions in linear Paul traps \cite{Monroe2021RMP,Richerme2014, Jurcevic2014, Brydges2019} or in 2d Penning traps \cite{Bohnet2016}, and atoms in cavities \cite{Lietal2023}; and they could be readily tested with state-of-the-art experiments.

\emph{Acknowledgements.} We gratefully acknowledge useful suggestions and comments from M. Fagotti, G. Pupillo, C. L. Sriram, and R. \'Swi\k{e}tek. This project is supported by ANR (22-PETQ-0004, projet QubitAF). No AI tool has been used for this scientific project nor for the writing of the manuscript.

\newpage

{\bf Methods} 

\medskip

\begin{figure}[ht!]
\begin{center}
\includegraphics[width=0.8\columnwidth]{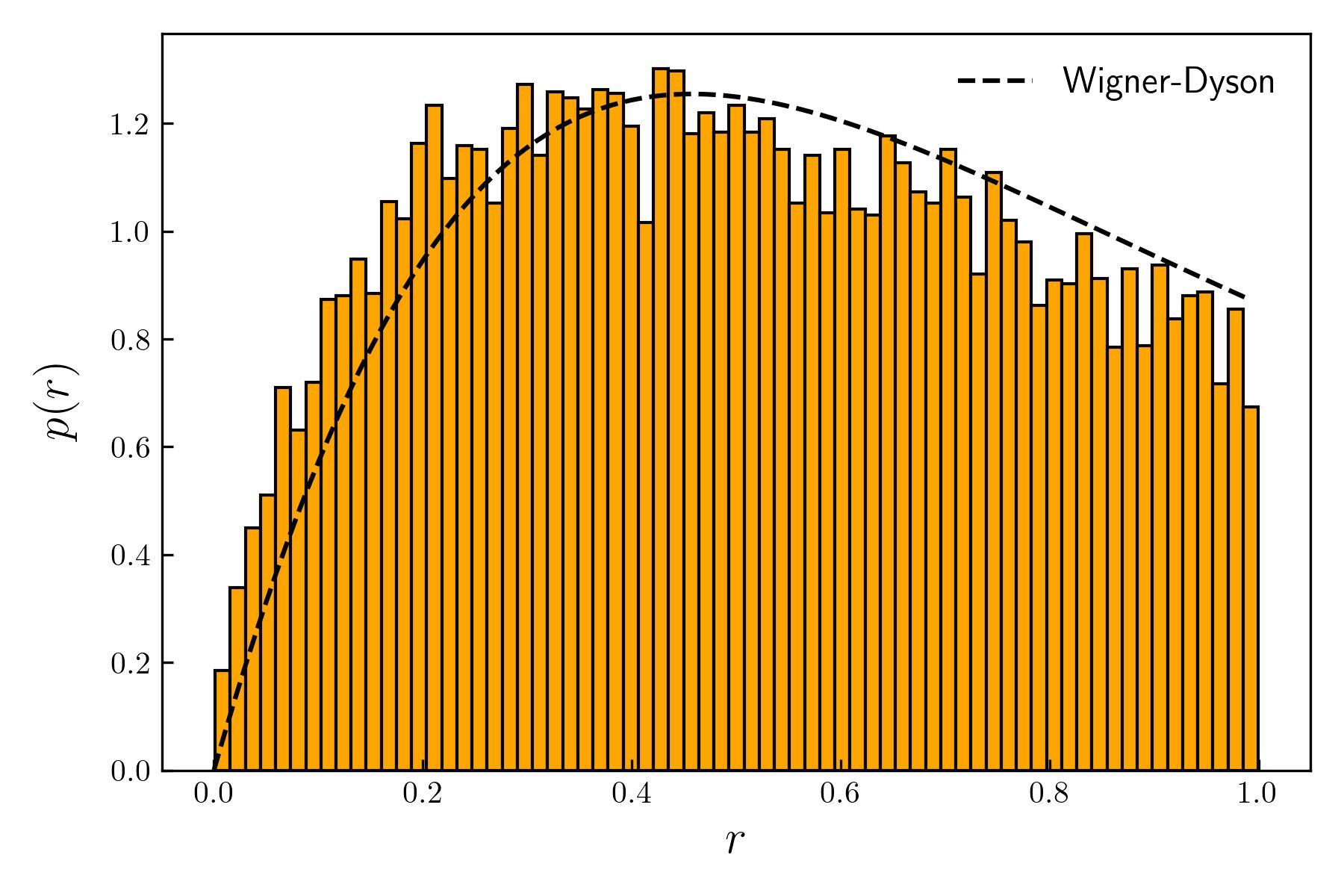}
\caption{\emph{Level statistics of the 1d LR XX model.} Probability of the ratio between consecutive level spacings, $r = \min(r_n, r_{n-1})$ with $r_n = (E_{n+1}-E_n)/(E_n-E_{n-1})$, for the 1d LR XX model with $\alpha=0.5$ and a field $\Omega/{\cal J} = 0.3$. The data refers to the spectrum for $N=22$ in the Hilbert-space sector with zero momentum and symmetry under lattice reflection and spin inversion (rotation of $\pi$ around $x$). The dashed curve shows the prediction for Wigner-Dyson statistics, $p_{\rm WD}(r) = (8/27) (r+r^2)/(1+r+r^2)^{5/2}$ \cite{Atasetal2013}.}
\label{f.level}
\end{center}
\end{figure}

\emph{Additional data for the long-range XX model.} 
Fig.~\ref{f.level} shows the level statistics for the 1d LR XX model in a field. We consider levels in the sector of the Hilbert space with zero momentum and positive symmetry under lattice reflection and spin inversion. This is the sector for which we also examine the distribution of EEVs in all the examples of this work, and to which dynamics is confined when starting from the coherent spin state $|{\rm CSS}_{-x}\rangle$. We observe that the statistics is compatible with the existence of level repulsion, as expected in chaotic quantum systems, and it approaches the statistics expected from the Wigner-Dyson prediction for random matrices in the Gaussian orthogonal ensemble \cite{Atasetal2013}. Therefore this indicator of quantum chaos cannot detect the failure of LR systems to thermalize \cite{Leroseetal2025}. 

\begin{figure}[ht!]
\begin{center}
\includegraphics[width=\columnwidth]{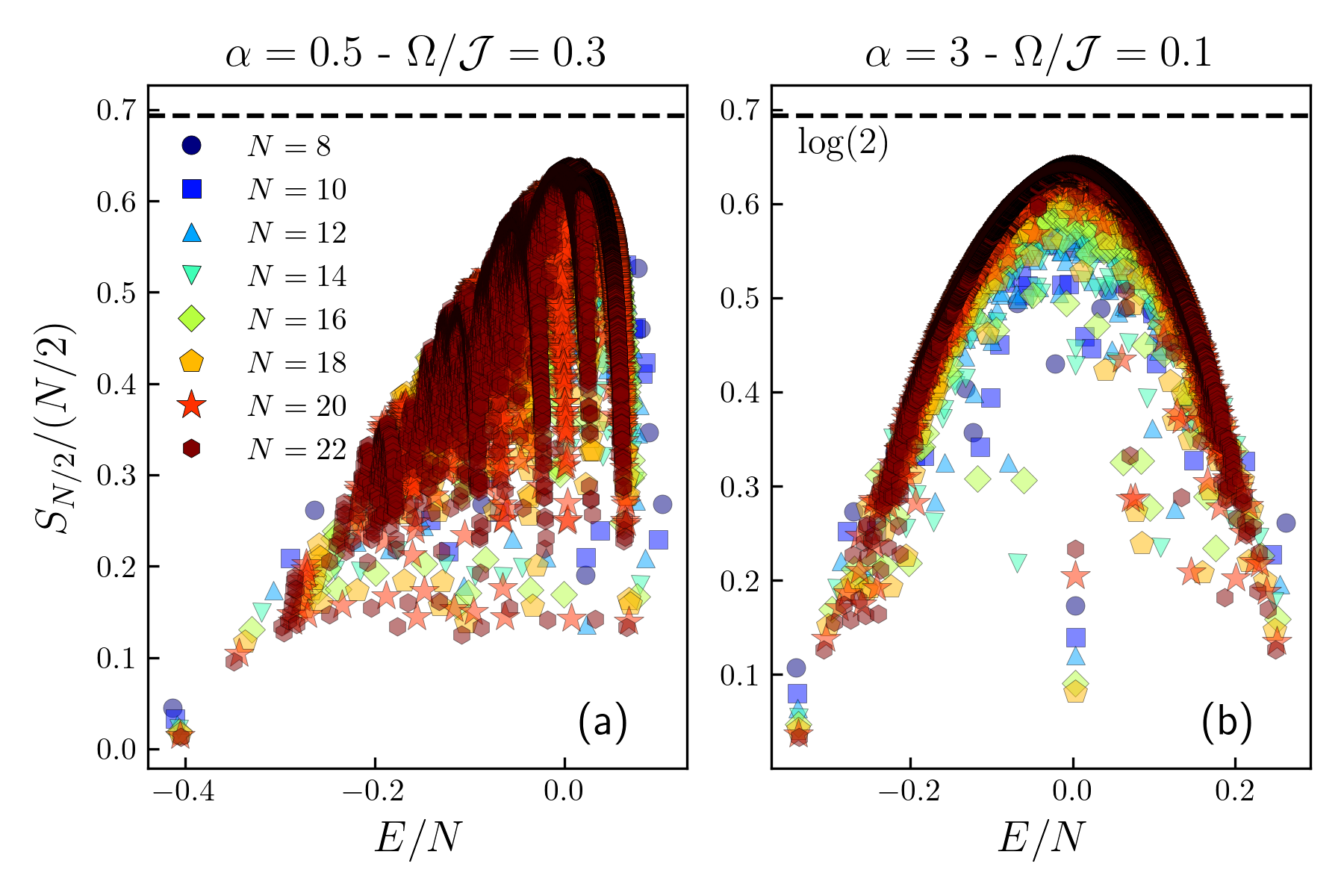}
\caption{\emph{Entanglement entropies (EEs) of eigenstates for long-range vs. short-range interactions.} (a) Half-system EEs (per spin) for the 1d LR XX model with $\alpha=0.5$ and a field $\Omega/{\cal J} = 0.3$. The dashed line marks the upper bound $\log 2$; (b) same quantity for the 1d XX model with $\alpha=3$ and $\Omega/{\cal J} = 0.1$.}
\label{f.entropy}
\end{center}
\end{figure}

Fig.~\ref{f.entropy}(a) shows the half-system entanglement entropy for the 1d LR XX model. We observe that the atypical nature of many eigenstates manifests itself in anomalously low entanglement entropies compared to the maximal one in each energy sector, which appears to follow a volume-law scaling with an entropy per spin which comes close to the maximum entropy of $\log 2$ (expected for states at zero energy, populated thermodynamically at infinite temperature). 

\begin{figure*}[ht!]
\begin{center}
\includegraphics[width=0.7\textwidth]{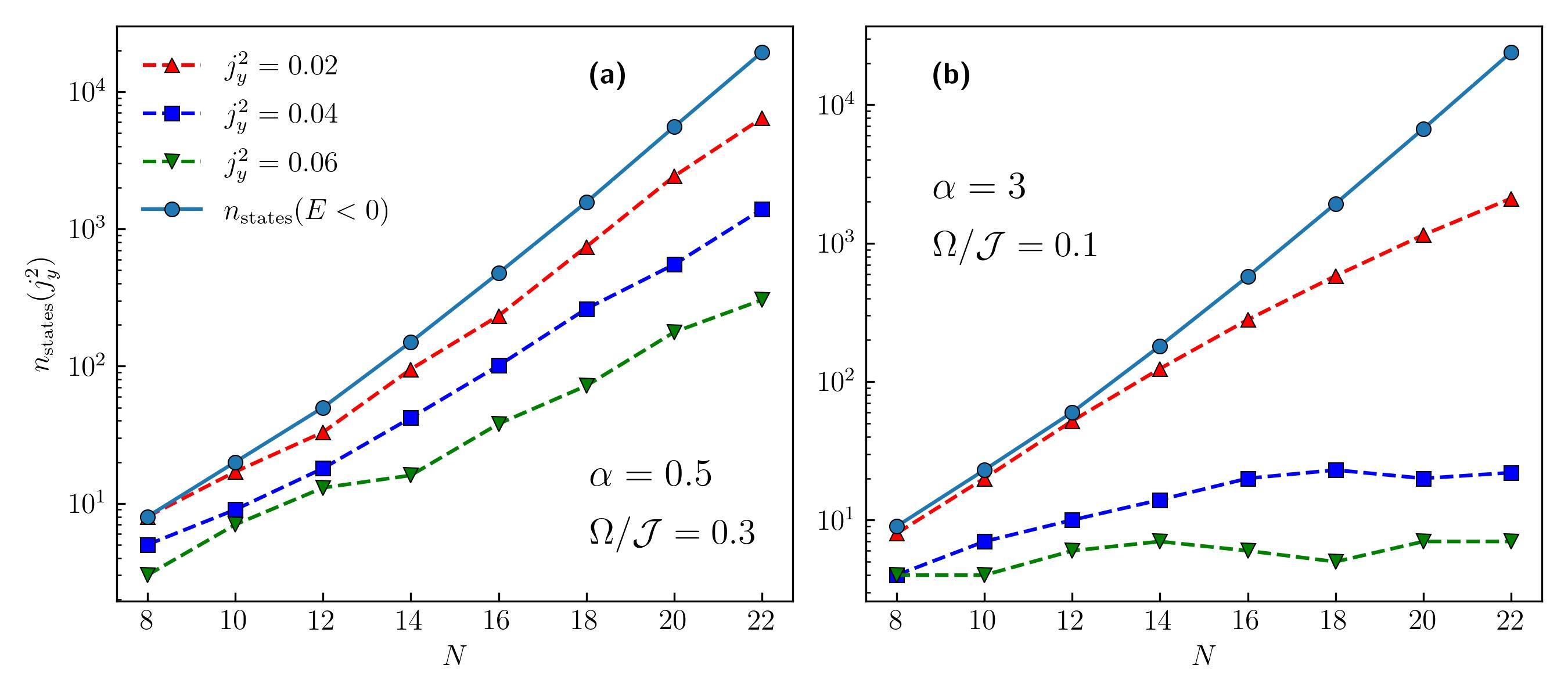}
\caption{\emph{Scaling of the number of atypical states.} (a) Number of negative-energy eigenstates $n_{\rm states}(j_y^2)$ with $\langle (J^y)^2/N^2 \rangle_{aa} > j_y^2$ as a function of system size for the 1d LR XX model with $\alpha = 0.5$ and $\Omega/{\cal J} = 0.3$, compared to the total number of states in the sector with zero-momentum and positive symmetry under lattice reflection and spin inversion. (a) Same quantities for the short-range chain with $\alpha = 3$ and $\Omega/{\cal J} = 0.1$.}
\label{f.exponentials}
\end{center}
\end{figure*}

Fig.~\ref{f.exponentials} offers an estimate of the scaling of the number of atypical states with system size. We identify the atypical states with those who have an anomalously large value of the fluctuations of the $J^y$ collective spin component, $j_y^2 = \langle (J^y)^2/N^2 \rangle$, associated with anomalously large correlations for the $S^y$ spin components. Long-range correlations in an atypical eigenstate are associated with $j_y^2$ not decreasing with system size, and hence with $m_y^2 \neq 0$ (see main text). On the other hand, states complying with the thermodynamic behavior in the paramagnetic phase would have $j_y^2 \sim O(1/N)$. We count the number of states with negative energy (i.e., associated with the physics of the ferromagnetic 1d LR XX model) whose EEV  $\langle (J^y)^2/N^2 \rangle_{aa}$ exceeds a given value $j_y^2$, and we indicate it as $n_{\rm states}(j_y^2)$.  Fig.~\ref{f.exponentials}(a) shows that, for the 1d LR XX model with $\alpha = 0.5$ and $\Omega/{\cal J} = 0.3$, this number appears to scale \emph{exponentially} with system size for various thresholds $j_y^2$. This suggests that the number of atypical states exhibiting anomalously large correlations grows exponentially with system size; yet, the exponential growth is still lower than that of the total number of states at negative energy (in the same symmetry sector), $n_s(E<0)$. Hence the atypical states would strictly speaking tend towards a set of zero measure in Hilbert space, so that the MC averages reproduce the expected paramagnetic behavior of the system.  Nonetheless they form an exponentially large set, and they are sufficiently numerous to control the scaling of the fluctuations of the EEVs around the MC average. 

This behavior is to be contrasted with the case of short-range interactions ($\alpha = 3, \Omega/{\cal J} = 0.1$) shown in Fig.~\ref{f.exponentials}(b). There the number of states with anomalously large correlations appears to not scale with system size beyond a given size, dependent on the threshold $j_y^2$. 

\begin{figure*}[ht!]
\begin{center}
\includegraphics[width=0.84\textwidth]{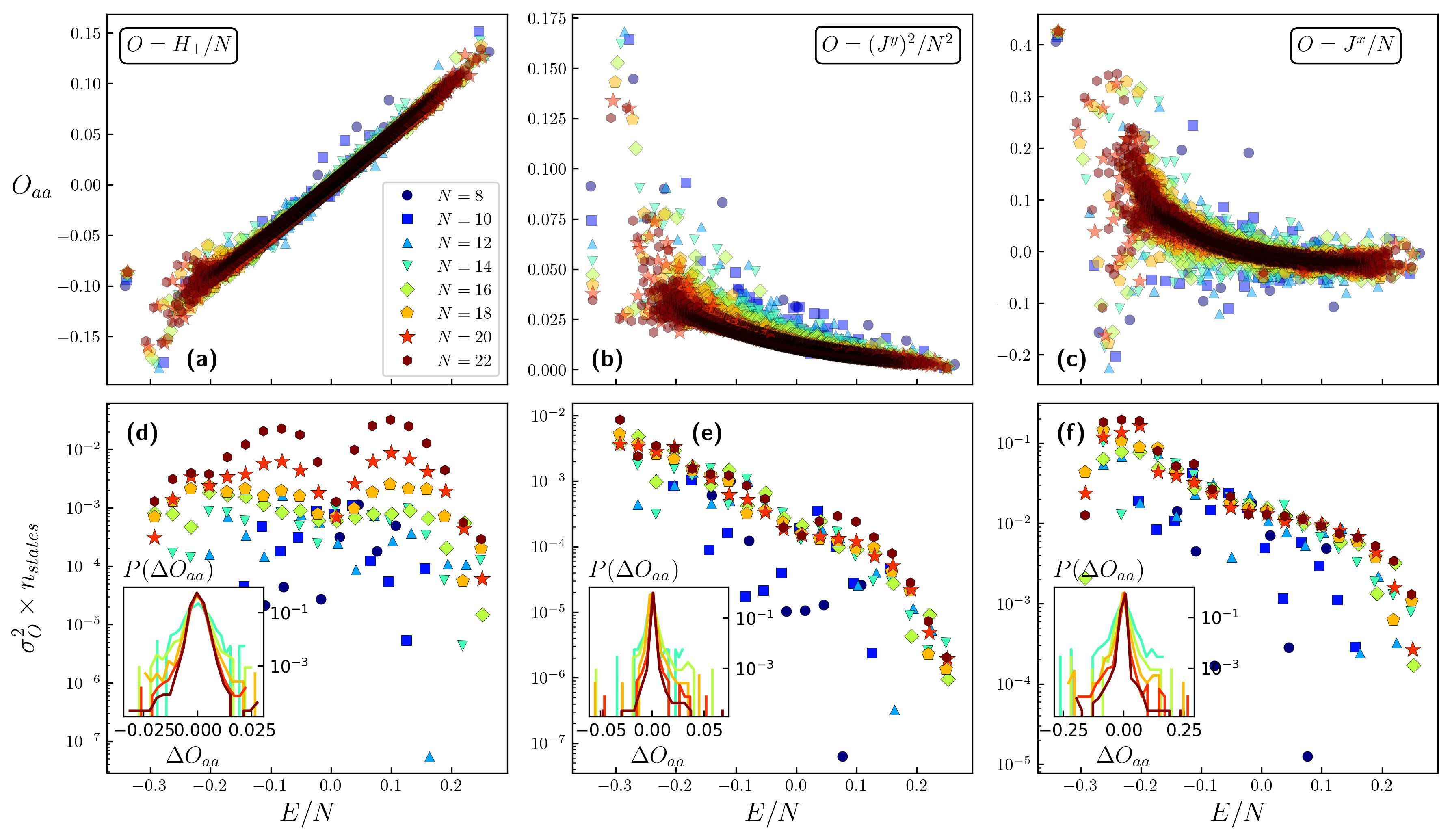}
\caption{\emph{Typicality of the eigenstates with short-range interactions.} (a, b, c) Eigenstate expectation values (EEV) $O_aa$ for three different observables, $O = H_{\perp}/N$, $(J^y)^2/N^2$ and $J^x/N$, for 8 different sizes, obtained for the 1d XX model with $\alpha=3$ and a field $\Omega/{\cal J} = 0.1$. (d, e, f) Scaling behavior of the mean squared deviations of EEVs, $\Delta O_{aa}$, from their local microcanonical average, for the three above observables. Insets: distribution of the local deviations of EEVs, $\Delta O_{aa}$, evaluated over the states with negative energy.}
\label{f.typicality}
\end{center}
\end{figure*}

\emph{Compliance of short-range interacting models with ETH.} The atypicality of eigenstates for long-range interactions ($\alpha = 0.5$ in 1d) is in stark contrast with the case 
of short-range interactions, $\alpha=3$, shown in Fig.~\ref{f.typicality}. There we observe that EEVs appear to accumulate over increasingly smooth curves as the system size increases \cite{dalessio_quantum_2016} (Fig.~\ref{f.typicality}(a-c)). The deviations of EEVs from local MC averages appear to decay exponentially with system size (in compliance with ETH), and are compatible with the scaling $\sigma_O^2 \sim 1/n_{\rm states}$, where $n_{\rm states}$ is the number of states in energy windows of width $W/20$ (with $W$ the full width of the spectrum) over which the averaged squared deviations $\sigma_O^2$ are also calculated (see main text). This is shown by the collapse of $\sigma_O^2$ data for different sizes when multiplied by  $n_{\rm states}$, as seen in Fig.~\ref{f.typicality}(e,f) for the observables $(J^y)^2/N^2$ and $J^x/N$. On the other hand, for $H_{\perp}/N$ this scaling appears only for energies $E/N \approx 0$. This same scaling has been found in Ref.~\cite{Beugelingetal2014} in models complying with ETH.  
Gradual collapse onto a smooth curve is also observed for the half-system entanglement entropies, shown in Fig.~\ref{f.entropy}(b).

\bibliography{squeezing.bib}

\begin{thebibliography}{46}
\bibitem[{SM()}]{SM}
\bibinfo{note}{See {S}upplemental {M}aterial ({SM}) for the density of states
  of the 1d LR XX model; additional data about atypicality of its eigenstates;
  and for details about rotor+spin-wave theory.}
\end{thebibliography}

\null 
\newpage

\begin{center}
{\bf SUPPLEMENTAL  MATERIAL}

\bigskip

\noindent
{\bf \emph{Breakdown of thermalization in quantum spin systems with long-range interactions}}

\end{center}

\bigskip

\section{Density of states of the 1d long-range XX model}  

\medskip

\begin{figure}[ht!]
\begin{center}
\includegraphics[width=\columnwidth]{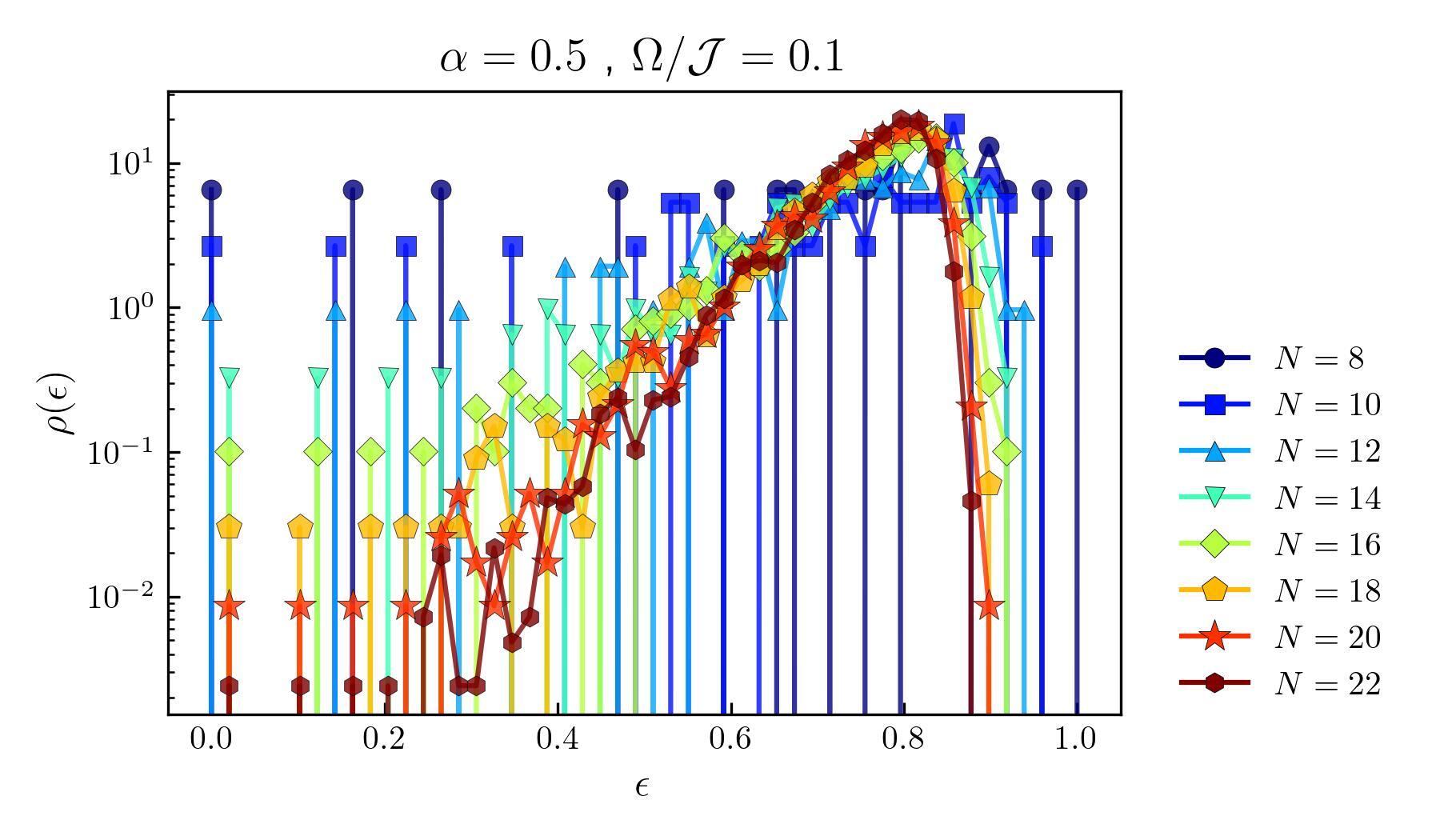}
\includegraphics[width=\columnwidth]{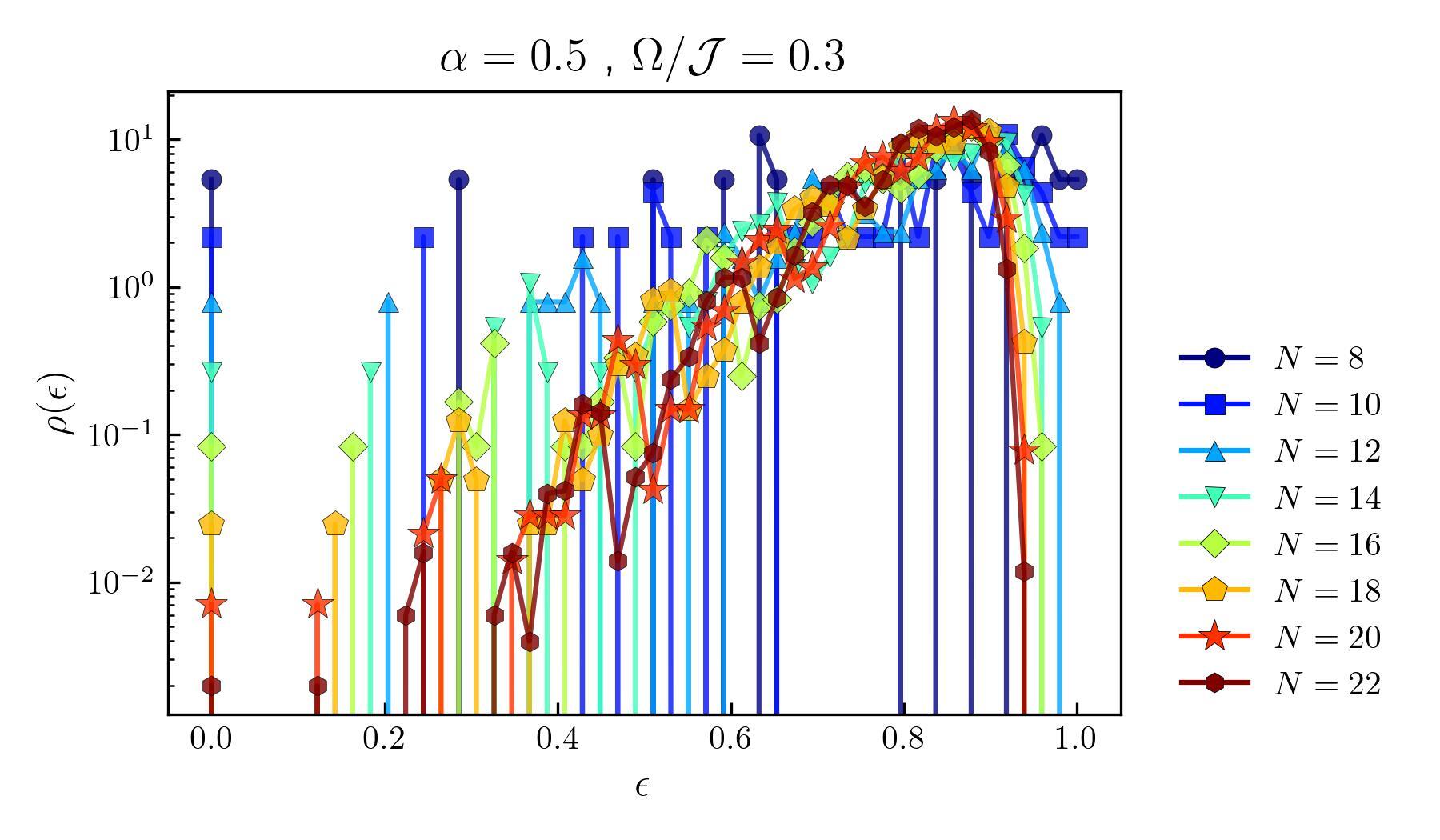}
\includegraphics[width=\columnwidth]{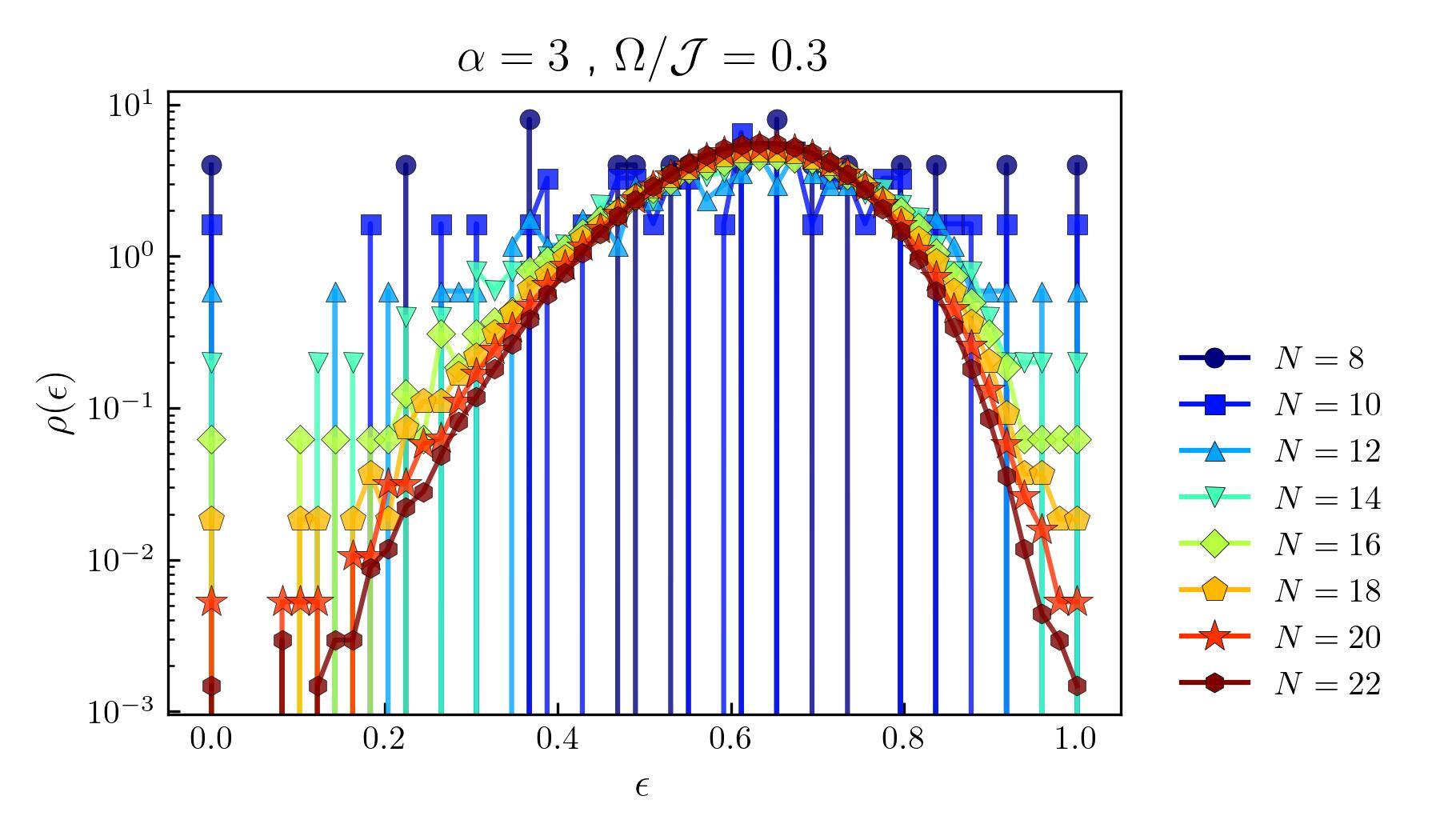}
\caption{\emph{Density of states for the 1d LR XX model.} Upper panel: $\alpha=0.5$ and $\Omega/{\cal J} = 0.1$; middle panel: $\alpha=0.5$ and $\Omega/{\cal J} = 0.3$; lower panel: case with short-range interactions, $\alpha=3$ and $\Omega/{\cal J} = 0.3$. We only consider states in the zero momentum sector, and with positive symmetry under lattice reflection and spin inversion.}
\label{f.DOS}
\end{center}
\end{figure}

{Previous works on quantum systems with long-range interactions \cite{Russomannoetal2021,PalS2025, Srirametal2026} have observed that the spectrum of a finite-size system appears to be structured into energy bands, which collapse to highly degenerate energy levels in the limit of all-to-all interactions. The persistence of a band-like structure of the energy spectrum for arbitrary sizes would imply the inequivalence between the microcanonical and the canonical ensemble, since the microcanonical entropy (related to the logarithm of the density of states) would no longer be a monotonic function of the energy (for negative energies).

In the case of the 1d LR XX model ($\alpha=0.5$) we consider, the apparent organization of the spectrum into bands disappears for sufficiently big sizes, so that ensemble equivalence is expected to emerge in the thermodynamic limit. This is shown in Fig.~\ref{f.DOS}, where the density of states $\rho(\epsilon) = n(E,E+\delta E)/(n_{\rm tot} ~d\epsilon$ is plotted as a function of the reduced energy $\epsilon = E - E_{\rm min}/(E_{\rm max} - E_{\rm min})$. Here $E_{\rm min}$ and $E_{\rm max}$ are the extremal energy eigenvalues, $n_{\rm tot}$ is the total number of states (in the zero momentum sector, and with positive symmetry under lattice reflection and spin inversion); $\delta E = (E_{\rm max}-E_{\rm min})/50$ is the width of each energy bin, and $d\epsilon = 1/50$ the same width in reduced units. Clearly, as the size increases, the density of states becomes an increasingly smooth function of the energy, and monotonically increasing up to zero energy (which does not correspond to the center of the spectrum, since the spectrum is not symmetric). A similar smoothening with increasing system size is also observed in the case of short-range interactions ($\alpha=3$), also shown in Fig.~\ref{f.DOS}.}

{In light of this result, we conjecture that a similar behavior (smoothening of the density of states and disappearance of energy bands upon increasing the system size) is exhibited for any $\alpha>0$, but the size required to see energy bands disappear is larger the smaller $\alpha$.}

\bigskip

\begin{figure*}[ht!]
\begin{center}
\includegraphics[width=0.95\textwidth]{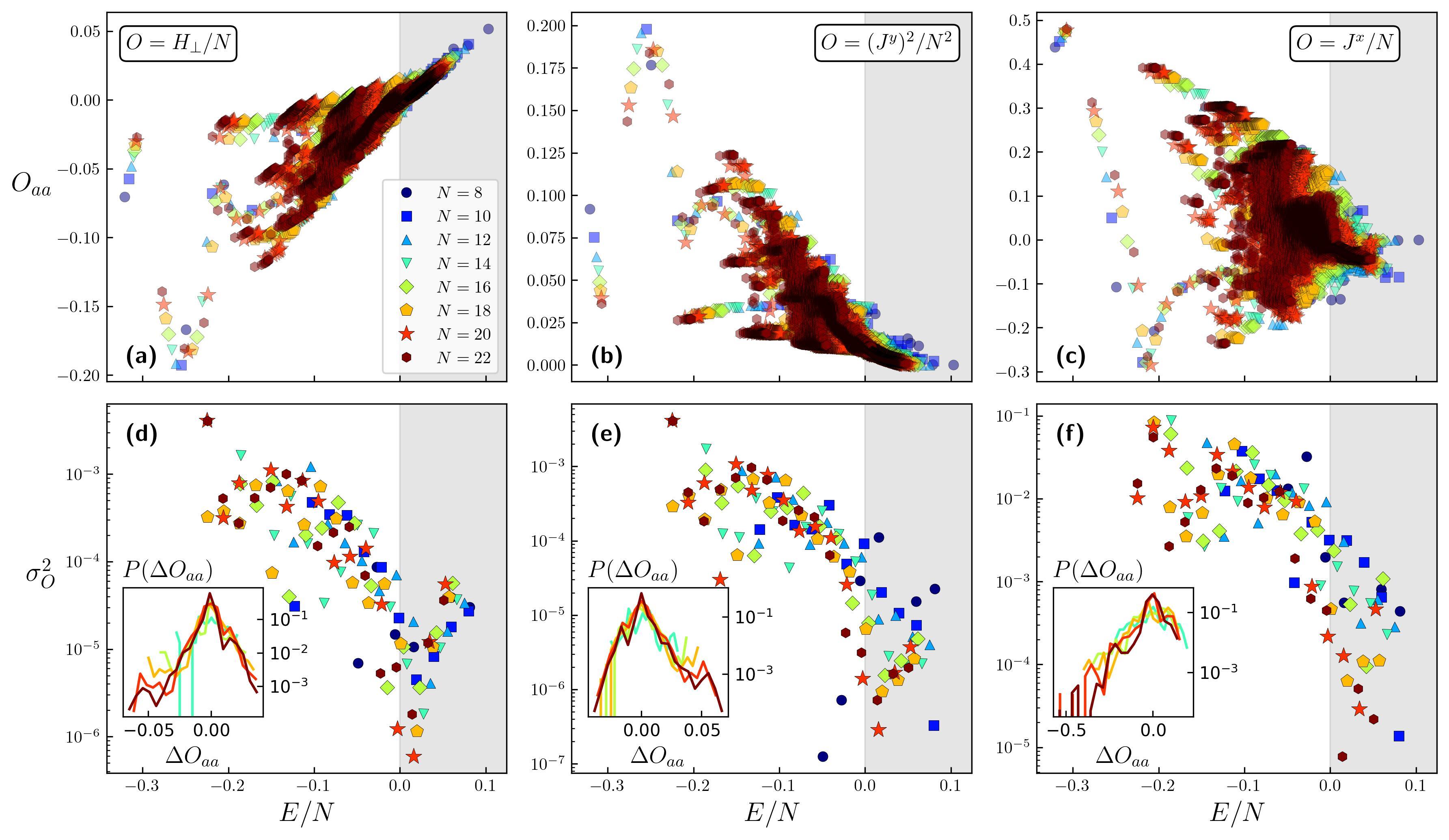}
\caption{\emph{Atypicality of the eigenstates with LR interactions.} (a, b, c) Eigenstate expectation values (EEV) $O_{aa}$ for three different observables, $O = H_{\perp}/N$, $(J^y)^2/N^2$ and $J^x/N$, for 8 different sizes, obtained for the 1d LR XX model with $\alpha=0.5$ and a field $\Omega/{\cal J} = 0.1$.  (d, e, f) Scaling behavior of the mean squared deviations of EEVs, $\Delta O_{aa}$, from their local microcanonical average, for the three above observables. Insets: distribution of the local deviations of EEVs, $\Delta O_{aa}$, evaluated over the states with negative energy.}
\label{f.atypicality2}
\end{center}
\end{figure*}

\section{Additional data for the atypicality of eigenstates of a long-range interacting Hamiltonian}  

\medskip

Fig.~\ref{f.atypicality2} shows the atypical behavior of EEVs for the 1d LR XX model with field  $\Omega/{\cal H} = 0.1$. In particular the absence of scaling of the variances $\sigma_O^2$ of EEVs around their local microcanonical averages, in violation of ETH, is evident for all considered observables for negative energies $E/N<0$.

\bigskip

\section{Robustness of rotor-spin-wave separation with long-range interactions}  

\medskip

The main text presents results obtained with rotor+spin-wave (RSW) theory \cite{Roscildeetal2023b,Roscildeetal2023}, applied to the dynamics  of the 1d LR XX model in a field along $x$, initialized from a coherent spin state opposite to the field. This theory is based on the approximate separation between the dynamics occurring in the Hilbert-space sector of maximal spin length, ${\bm J}^2 = N/2(N/2+1)$ (akin to the dynamics of a planar rotor in a field), and the dynamics occurring outside of that sector, which is approximated as harmonic spin-wave dynamics. This separation is justified so long as the spin waves remain dilute, namely if the dynamics remains confined close to the maximal-spin sector. 

\begin{figure}[ht!]
\begin{center}
\includegraphics[width=0.9\columnwidth]{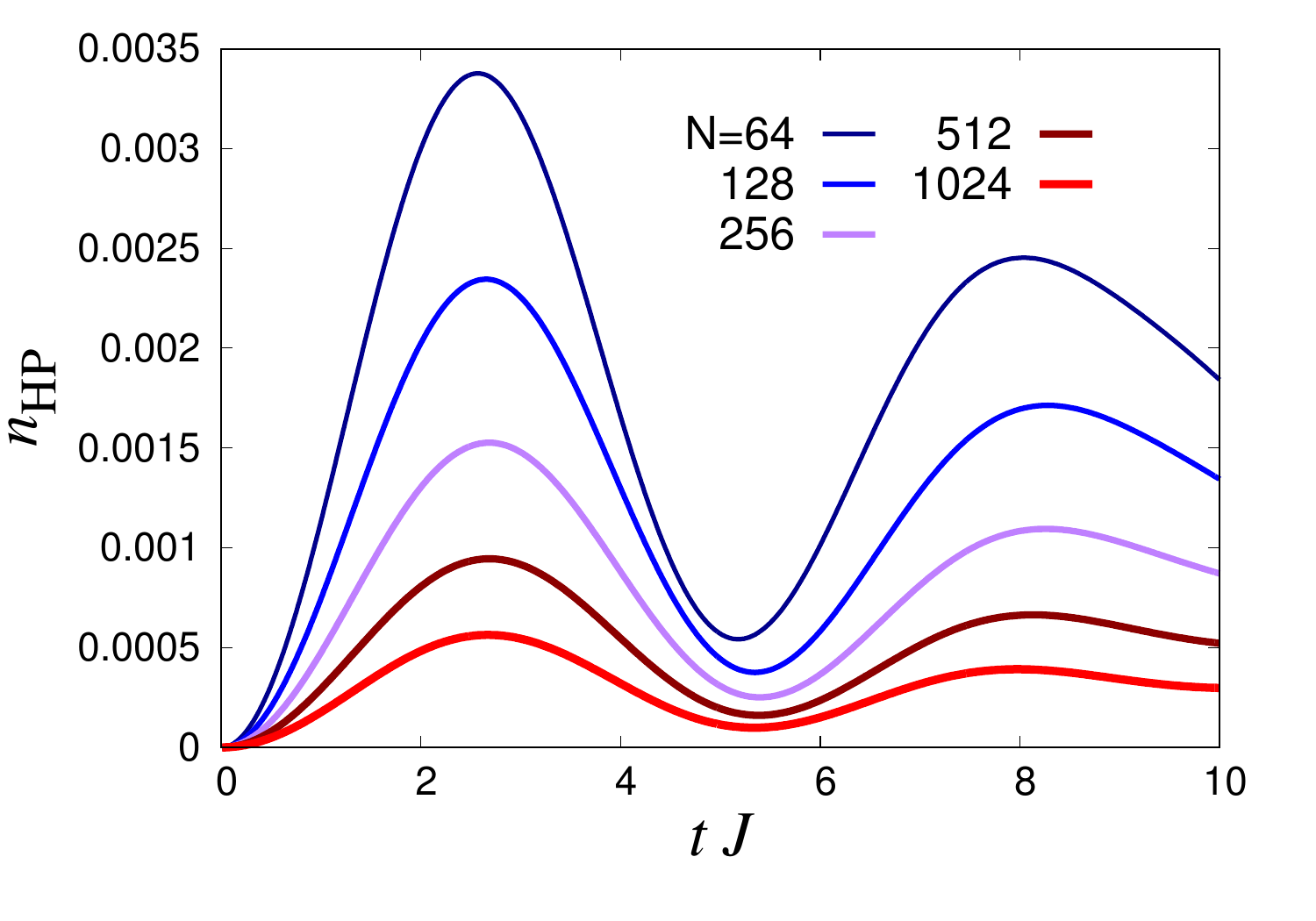}
\caption{Density of HP bosons at finite momentum $n_{\rm HP}$ as a function of time in the quench dynamics of the 1d LR XX model with $\alpha=0.5$ and $\Omega/{\cal J} = 0.3$.}
\label{f.HP}
\end{center}
\end{figure}

In practice the Hamiltonian of the system is approximated as \cite{Roscilde2025}
\begin{eqnarray}
H &\approx& \frac{(K^z)^2}{2I} - \Omega K^x  \\
&+& \sum_{k \neq 0} \left [ A_k b_k^\dagger b_k + \frac{B_k}{2} \left ( b_k b_{-k} + {\rm h.c.} \right) \right] + {\rm const.} \nonumber 
\end{eqnarray}
Here ${\bm K} = (K^x, K^y, K^z)$ is the collective spin operator projected onto the sector of maximal spin length, spanned by Dicke states, and describing the rotor degree of freedom. The operators $b_k, b_k^\dagger$ are related to Holstein-Primakoff (HP) bosons at wavevector $k = 2\pi p/N$ with $p = 1,..., N-1$, describing harmonic spin waves. Their vacuum corresponds to the initial state of the dynamics, namely the $|{\rm CSS}_{-x}\rangle$ state. The inverse moment of inertia for the rotor is 
\begin{equation}
\frac{1}{2I} = \frac{\cal J}{(N-1)K_{\alpha}} \sum_{r\neq 0} \frac{1}{r^\alpha}
\end{equation}
while the spin-wave energy cost $A_k$  and pair-production/destruction rate $B_k$ take values
\begin{align}
A_k =  &~ {\cal J} \left ( 1- \frac{\gamma_k}{2 K_\alpha} \right ) - \Omega \\
B_k = &  - {\cal J} \frac{\gamma_k}{2 K_\alpha} 
\end{align}
where 
\begin{equation}
\gamma_k = \sum_{r \neq 0} \frac{e^{ikr}}{r^\alpha} 
\end{equation}
with $\gamma_0 = K_\alpha$.

Unlike for ordinary spin-wave theory, for the dynamics of interest the initial state is not the mean-field ground state of the Hamiltonian, and therefore spin-wave excitations can develop instabilities. Such instabilities would translate into the fact that spin-wave eigenfrequencies $\omega_k = \sqrt{A_k^2 - B_k^2}$ can become imaginary for some wavevectors. This would imply their exponential proliferation, and hence the breakdown of the conditions for separation between rotor and spin-wave dynamics. Yet, as shown in Ref.~\cite{Roscilde2025}, for long-range interactions, $\alpha<d$, spin waves built around the state opposite to the field are stable (namely all of their frequencies are real) for all system sizes up to a critical field, which, for the model of interest, is considerably larger than the largest field we considered ($\Omega/{\cal J} = 0.3$). 

In fact it is easy to realize that RSW theory is exceptionally well justified for the model and dynamics of interest. Starting from the vacuum of HP bosons, the above Hamiltonian generates pairs of bosons at opposite wavevectors at a rate $B_k$, but with an energy cost of $2 A_k$ for each pair, which keeps the number of bosons limited. 
Moreover, $\gamma_k \sim k^{d-\alpha}$ for $k \to 0$ (as Fourier transform of the $r^{-\alpha}$ function in $d$ dimensions), namely it is finite for every finite $k$. Hence, since $K_\alpha \sim N^{1-\alpha/d}$, then one has that $B_k \sim N^{-1+\alpha/d}$, namely the pair-production rate for HP bosons decreases with increasing size. On the other hand, $A_k \to {\cal J} - \Omega$ when $N\to \infty$, hence it remains finite and it becomes size-independent.  As a result, the pair production becomes less effective the larger the system size, producing a decreasing density of HP bosons when size increases. This is clearly shown in Fig.~\ref{f.HP}, where the density of the HP bosons generated dynamically, $n_{\rm HP} = N^{-1} \sum_{k \neq 0} \langle b_k^\dagger b_k \rangle$, appears to decrease with system size.  This means that the assumption behind RSW theory, i.e. the decoupling between rotor and spin-wave dynamics, is increasingly well justified with increasing system size, since the leakage of the dynamics away from the Dicke sector is less and less pronounced.

\end{document}